\documentstyle[12pt]{article}
\input {epsf}

\newcommand{\beq}{\begin{equation}}
\newcommand{\eeq}{\end{equation}}
\newcommand{\beqa}{\begin{eqnarray}}
\newcommand{\eeqa}{\end{eqnarray}}

\def\openone{\leavevmode\hbox{\small1\kern-3.8pt\normalsize1}}
\def\ket#1{|\,#1\,\rangle}
\def\bra#1{\langle\, #1\,|}
\def\braket#1#2{\langle\, #1\,|\,#2\,\rangle}

\begin{document}

\title{Intermediate states in quantum cryptography and Bell inequalities}
\author{
H. Bechmann-Pasquinucci$^1$ and N. Gisin$^2$
\\
\small{ $^1${\rm UCCI.IT}, \it via Olmo 26, I-23888 Rovagnate (LC), Italy} 
\\
\small{\it $^2$Group of Applied Physics, University of Geneva, CH-1211, 
Geneva 
4,Switzerland}}
\date{August 6, 2002}
\maketitle
\abstract{Intermediate states are known from intercept/resend 
eavesdropping in the BB84 quantum cryptographic protocol. But they also 
play fundamental roles in the optimal eavesdropping strategy on BB84 and 
in the 
CHSH inequality. We generalize the intermediate states to arbitrary 
dimension and consider intercept/resend eavesdropping, optimal 
eavesdropping on the generalized BB84 protocol and present a generalized 
CHSH inequality for two entangled quNits based on these states.}

\section{Introduction}
The quantum cryptographic protocol, known as the BB84 \cite{BB84}, was 
originally 
developed for qubits. In this protocol the legitimate users, Alice and 
Bob, both use the same two mutually unbiased 
bases $A$ 
and $A'$. Alice use them for state preparation\footnote{Notice that Alice 
may 
use a maximally entangled state of two qubits for preparing the state she 
sends to 
Bob, since a measurement on one qubit will 'prepare' the state of the 
other qubit.} and Bob chooses between the two bases for his measurement. 
But 
an eavesdropper performing the simple intercept/resend eavesdropping, 
may chose to measure in what is known as the intermediate basis or the 
Breidbart basis \cite{exp}. In the case of qubits it is possible to form 
four 
intermediate states, which falls into two mutually unbiased bases. However 
the eavesdropper need only use one of these bases. 

It turns out that it is not only in the simple intercept/resend 
eavesdropping that these intermediate states appear. Also in the optimal 
eavesdropping strategy \cite{DC,CBKG}, which consists of the eavesdropper 
using the 
optimal cloning machine, these states enters. In this case, they appear at 
the point where Bob and the eavesdropper, Eve, have the same amount of 
information, i.e. where their information lines cross. At this point their 
mixed states may be decomposed into a mixture of some of the intermediate 
states. 

That the intermediate states also appear in the optimal eavesdropping 
strategy, also explains a curious observation. Namely, that the amount of 
information obtained by the eavesdropper at the crossing point 
between the information lines using 
optimal eavesdropping, and the amount of information she obtains 
performing 
intercept/resend eavesdropping in the intermediate basis, is the same. 
However, the error rates are quite different.

Further more intermediate states reappear in the 
Clauser-Horne-Shimony-Holt (CHSH) inequality \cite{CHSH} for
two entangled qubits. Where the maximal violation is obtained when on the
first qubit the measurement settings correspond to the two mutually
unbiased bases $A$ and $A'$, and on the second qubit the two intermediate
bases. Moreover when introducing the same kind of noise as the 
eavesdropper does in the optimal eavesdropping strategy, the Bell 
violation naturally  decreases. But it is interesting to notice that 
for the critical disturbance where the classical limit is reached,   
Bob and 
Eve have the same amount of information, i.e. this happens at the crossing 
point of the 
information lines. 
This crossing point between the two information lines is a very important 
point, since upto this limit Alice and Bob can use the fact that they have 
more mutual information than the eavesdropper and they can create a secure 
key just by using classical error correction and one-way privacy 
amplification. 
Hence the CHSH inequality for qubits can be used as a security measure 
\cite{fuchs97,qcrmp}.

In the three situation just described, intercept/resend 
eavesdropping, optimal eavesdropping and the CHSH-inequality, the 
intermediate states keep reappearing and seem to play a fundamental role.
 
A natural question to ask is 'what happens in higher dimensions?'. 
This is the question we try  to answer, at least partially, here. 
It 
is possible to generalize 
the BB84 protocol to arbitrary dimension \cite{BT,BP,Sw,DC,CBKG}, simply 
by adding basis vectors 
to the two mutually unbiased bases, so that for $N$ dimension each basis 
contain $N$ vectors. The intermediate states may also be generalized to 
arbitrary dimensions. However, in higher dimensions they do in general not 
form bases. But it is possible to associate with each intermediate state 
a projector, which represents  a binary measurement. 

With the use of these generalized intermediate states we investigate 
intercept/resend eavesdropping, optimal eavesdropping and a generalized 
CHSH inequality in arbitrary dimension to see if they play the same role 
as in two dimension. 

In section 2 we introduce the intermediate states for quNits. In section 3 
we shortly 
discuss intercept/resend eavesdropping using the intermediate states. In 
section 4 we compare optimal eavesdropping with the intercept/resend 
eavesdropping strategy. Then in section 5 we present a generalized Bell 
inequality for two entangled quNits. In section 6 we consider the Bell 
violation as a function of the disturbance the optimal eavesdropping 
strategy would lead to. The last sections of the paper is devoted to a 
study of the inequality we have presented. In section 7 we discuss some 
features of the inequality by   
giving examples in three dimensions. Since recently the 
strength of a 
Bell inequality has been measured in terms of its resistance to noise we 
discuss this issue in section 8. Section 9 is devoted to a brief study of the
required detection efficiency. Finally in section 10 we have 
conclusion and discussion.

\section{The intermediate states}
The quantum cryptographic protocol BB84 can easily be generalized to 
arbitrary dimension, this has already been discussed in the 
literature \cite{BT,Sw}. 
The protocol works in exactly the same way as for qubits with the sole 
exception that for quNits each of the two mutually unbiased bases $A$ and 
$A'$ used by Alice 
and Bob contain $N$ basis states instead of two. So that Alice sends at 
random (and with equal probability) one of the $2N$ possible states and 
Bob choses to measure in one of 
the two bases $A$ and $A'$. 

In this section we define the intermediate 
states between these two bases. The basis $A$ is chosen as the 
computational 
basis, 
\beq
\ket{{a}_{0}},  \cdots ,\ket{{a}_{N-1}},
\eeq
and the second basis, 
$A'$, is the Fourier transform of the computational basis:
\begin{equation}
\ket{{a}_{k}'}=\frac{1}{\sqrt{N}}\sum_{n=0}^{N-1} \exp\left(\right. 
\frac{2\pi i~kn}{N}\left.\right)\ket{{a}_{n}}
\end{equation}
These two bases are mutually unbiased, i.e.
\begin{equation}
\braket{{a}_{n}}{{a}_{k}'}=\frac{\exp\left(\right.\frac{2\pi 
i ~kn}{N}\left.\right)}{\sqrt{N}} 
\label{eq:unb}
\end{equation}
This  means that the distance between pairs of state from the two bases is 
$\cos(\theta)=1/\sqrt{N}$.

Having two states, it is possible to define a state which lies exactly in
between the two, which means that it has the same overlap with both states
and it is the state closest to the two original states which has this
property. The intermediate state are obtained by forming all possible 
pairs of the states from the two bases. They are shown in the table below:
\begin{center}
\begin{tabular}{|c||c|c|c|c|}\hline
 &${a}_{0}'$& ${a}_{1}'$&$\cdots$ &${a}_{N-1}'$ \\\hline\hline
${a}_{0}$& ${m}_{00}$&${m}_{01}$&$ \cdot $&${m}_{0,N-1}$\\\hline
${a}_{1}$& ${m}_{10}$&${m}_{11}$&$ \cdot $&${m}_{1,N-1}$\\\hline
$\vdots $& $\cdot$& $\cdot$& $\cdot$& $\cdot$\\\hline
${a}_{N-1}$& ${m}_{N-1,0}$&${m}_{N-1,1}$&$ \cdot$ &${m}_{N-1,N-1}$\\\hline
\end{tabular}
\end{center}
Explicitly the intermediate state between 
$\ket{{a}_{n}}$ and $\ket{{a}_{k}'}$ 
is defined in the following way
\begin{eqnarray}
\ket{{m}_{nk}}&=&\frac{1}{\sqrt{C}}\left[\exp\left(\frac{2\pi i 
~kn}{N}\right) 
\ket{{a}_{n}} + \ket{{a}_{k}'}\right]
\end{eqnarray}
where $C=2(1+1/\sqrt{N})$ is the normalization constant and the phase 
comes 
from the overlap between $\ket{{a}_{n}}$ and $\ket{{a}_{k}'}$, see 
eq.(\ref{eq:unb}). The index of the $m$-states are such that the first 
index always refers to the $A$ and the second to the  $A'$-basis. Since 
each basis 
contains $N$ states it is possible to form $N^2$ intermediate states, 
simply by forming all pairs of states from the two bases.

In general the intermediate state $\ket{{m}_{\alpha \beta}}$ between two 
arbitrary initial states 
$\ket{\alpha}$ and $\ket{\beta}$ is defined as 
\begin{eqnarray}
\ket{{m}_{\alpha \beta}}=\frac{\sqrt{\braket{\alpha}{\beta}}\ket{\alpha} 
+\sqrt{\braket{\beta}{\alpha}}\ket{\beta}}
{\sqrt{2}\sqrt{|{\braket{\alpha}{\beta}}|+|{\braket{\alpha}{\beta}}|^{2}}}
\end{eqnarray}
The intermediate states may be defined in complete generality for 
arbitrary initial states and any number of them. In this case the 
intermediate state is found by forming the mixture of all the initial 
states with equal weight, the eigenstate state with the largest 
eigenvalue of this mixture corresponds to the intermediate state. Naturally 
these definitions are equivalent and lead to the same intermediate state.

Considering the intermediate states leads to the following 
conditional probabilities
\begin{equation}
p({{m}_{nk}}|{{a}_{n}})=p({{m}_{nk}}|{{a}_{k}'})
=\frac{1+\frac{1}{\sqrt{N}}}{2}\equiv F
\label{eq:psuc}
\end{equation}
Notice that this definition indeed recover the formula for cosine of half 
the angle: $\cos(\theta/2)=\sqrt{\frac{1+cos(\theta)}{2}}$. This is why 
the 
states have been named intermediate states, since they indeed lie in 
between the the two original states. 

Whereas the probability for making an error is 
\begin{equation}
p({{m}_{nk}}|{{a}_{q}})=p({{m}_{nk}}|{{a}_{p}'})
=\frac{1-\frac{1}{\sqrt{N}}}{2(N-1)}\equiv \frac{D}{N-1}
\label{eq:perror}
\end{equation}

It is important to notice that the intermediates states in general not are 
orthogonal, indeed 
\begin{eqnarray}
\braket{{m}_{kl}}{{m}_{nm}}&=&\frac{1}{\sqrt{N}C}\left[ 
\sqrt{N}
{\delta}_{kn}\exp\left( {\frac{2\pi 
i}{N}(mn-lk)}\right)\right.
  \nonumber\\&+&\left.\sqrt{N}{\delta}_{lm} 
+\exp\left( {\frac{2\pi i}{N}(m-l)k}\right)+
\exp\left( {\frac{2\pi i}{N}(m-l)n}\right)
\right]
\end{eqnarray}
This means that the generalized intermediate states do in general not form 
bases as 
in the two dimensional case. But they can still be used as binary 
measurements, this is discussed in the next section.

\subsection{Intermediate states as binary measurements}
It has just been shown that in general the intermediate states 
$\ket{{m}_{kl}}$ are not 
orthogonal, and hence they do not form bases as in the two dimensional 
case. It is however possible to use the corresponding projectors, 
$\ket{{m}_{kl}}\bra{{m}_{kl}}$ as binary measurements.

Since the intermediate states are non-orthogonal, it means that the 
corresponding binary measurements are mutually incompatible. In other 
words, none of them can be measured together but they have to be measured 
one by one.
A binary measurement, has as the name indicates two possible outcomes, 0  
and 1. Where the zero outcome is {\it interpreted} as 'I guess the state 
was not $\ket{{m}_{kl}}$', and the '1' outcome is {\it interpreted} as 
'I guess the 
state was $\ket{{m}_{kl}}$'. However, the answers are statistical, in 
the sense that there is a certain probability for making the wrong 
identification. 

It should be mentioned that the $N^2$ intermediate states 
constitutes a generalized measurement namely a so called POVM. We have 
\begin{eqnarray}
\sum_{n,k=0}^{N-1}\frac{1}{N}\ket{{m}_{nk}}\bra{{m}_{nk}}=\openone
\end{eqnarray}
However, we do not make use of this in what follows.

\section{Intercept/resend eavesdropping}
Suppose that the eavesdropper, Eve, performs the simple intercept/resend 
eavesdropping. This means that she intercepts the particle send by Alice,
performs a measurement and according to the result
prepares a particle which she then sends to Bob. She may choose to measure
in the same bases as Alice and Bob, but she may also choose to use the
intermediate states. In higher dimensions where the intermediate states do
not form bases, this strategy becomes a bit artificial. It is nevertheless 
interesting to consider it briefly.

In arbitrary dimension where the intermediate states corresponds to binary 
measurements, the intercept/resend strategy using these measurements 
may look like this: When ever Eve obtains a '1', which means she can make 
a good guess of the state, she prepares a new state and sends it to Bob. 
Whereas in the cases where she gets a '0', which means she is unable to 
make a good guess, she does not send anything to Bob. 
In this way we are only considering the cases where Eve does obtain 
a useful answer. This strategy, of course, gives rise to a huge amount of 
losses 
and errors on Bobs side, but it is however interesting to evaluate the 
amount of information that Eve obtains in this case, i.e. considering only 
the measurements where she gets a positive answer. 

The probability of making the correct identification is given by 
eq.(~\ref{eq:psuc}) and is equal to 
$\frac{1}{2}+\frac{1}{2\sqrt{N}}$. 
Whereas the probability of wrong identification, i.e. of an error is given 
by eq.(~\ref{eq:perror}) and is equal to 
$\frac{1}{(N-1)}(\frac{1}{2}-\frac{1}{2\sqrt{N}})$. This means that the 
(Shannon) information 
obtained  by Eve is given by \cite{Sw}
\begin{eqnarray}
{I}_{int,Eve}^{N}&=&\log_2(N)+\left(\frac{1}{2}+\frac{1}{2\sqrt{N}}\right)
\log_2 
\left(\frac{1}{2}+\frac{1}{2\sqrt{N}}\right)\nonumber\\&+&
\left(\frac{1}{2}-\frac{1}{2\sqrt{N}}\right)\log_2 
\left(\frac{1}{(N-1)}\left(\frac{1}{2}-\frac{1}{2\sqrt{N}}\right)\right)
\label{eq:infoint}
\end{eqnarray}
on the '1' outcomes of her measurements.

In the next section we will compare this amount of information to the 
amount of information
obtained performing optimal eavesdropping at the point where the 
information lines between Bob and Eve cross.

\section{The optimal cloning machine}
The optimal eavesdropping strategy in any dimension, is believed to be 
given by a asymmetric version of the quantum cloning machine \cite{Cerf} which 
clones 
optimally the two mutually unbiased bases \cite{CBKG}. Using this 
cloner, Eve 
can 
obtain two copies of different 
fidelity of the state  prepared by Alice. Usually Eve keeps the bad copy 
and sends the good one on to Bob. For a full description of how this 
eavesdropping strategy and of the cloning machine involved,  see 
\cite{CBKG}. 
Here we are only concerned with the final state that Bob receives, 
which means how the optimal eavesdropping strategy influence the state 
obtained by Bob. In the case of no eavesdropping Bob receives the same 
pure state as was send by Alice. But in the case of eavesdropping Bob 
receives a mixed state.

Assume that without eavesdropping Bob would have found the state 
$\ket{{a}_{n}}$ if measuring in the computational basis. The question is 
what happens to $\ket{{a}_{n}}$ as a result of the eavesdropping? Or in 
other words, how does the cloning machine influence the state 
$\ket{{a}_{n}}$? We are only interested in the final mixed state Bob 
receives, and that may be written as
\begin{equation}
\rho_B=F_B\ket{{a}_{n}}\bra{{a}_{n}}+\frac{D_B}{N-1}\sum_{j=0, ~j\neq 
n}^{N-1}
\ket{{a}_{j}}\bra{{a}_{j}}
\label{eq:rhobob}
\end{equation}
where $F_B$ is the fidelity and $D_B=1-F_B$ is the total disturbance. A 
similar 
expressing can be obtained for the $A'$ basis states.
As a result of the eavesdropping the amount of information that Bob gets 
is
\begin{eqnarray}
{I}_{opt,bob}^{N}=\log_2(N)+F_B\log_2(F_B)+(1-F_B)\log_2\left( 
\frac{1-F_B}{N-1}\right)
\end{eqnarray}

The optimal eavesdropping strategy is symmetric under the exchange of Bob 
and Eve. This means that the mixed state, ${\rho}_{E}$, which Eve 
obtains can be written on the same form as Bob's mixed state, just with 
different coefficients, i.e.
\begin{equation}
\rho_E=F_E\ket{{a}_{n}}\bra{{a}_{n}}+\frac{D_E}{N-1}\sum_{j=0, ~j\neq 
n}^{N-1}
\ket{{a}_{j}}\bra{{a}_{j}}
\end{equation}
And equivalently the amount of information obtained by Eve is given by
\begin{eqnarray}
{I}_{opt,eve}^{N}=\log_2(N)+F_E\log_2(F_E)+(1-F_E)\log_2\left(
\frac{1-F_E}{N-1}\right)
\end{eqnarray}

It is interesting and important to consider the point where the 
information lines between Bob and Eve cross. Since when Alice and Bob 
share more information that Alice and Eve, Alice and Bob can use one-way 
privacy amplification to obtain a secret key. Using the explicit form and 
coefficients of the cloning machine it is possible to show (this was done 
in \cite{CBKG}) that the information curves cross at the point where
\begin{eqnarray}
&&F_B=F_E=F=\frac{1}{2}+\frac{1}{2\sqrt{N}}\\
&&D_B=D_E=D=\frac{1}{2}-\frac{1}{2\sqrt{N}}
\end{eqnarray}
This is exactly the same fidelity (or probability of guessing correctly 
the state) that Eve obtained using the intercept/resend eavesdropping 
using the intermediate states. Which means that we have just shown  
that 
\begin{equation}
I_{Int,eve}^{N}={I}_{opt,eve}^{N}({\rm crossing~point})
\end{equation}

This is explained by the fact that, at the crossing point of 
the information lines, Eve's mixed state can be decomposed into 
a mixture of some of the intermediate states, namely
\begin{equation}
\rho_E^{cross}=\frac{1}{N}\sum_{j=0}^{N-1}
\ket{{m}_{nj}}\bra{{m}_{nj}}
\end{equation}
where again it has been assumed that $\ket{a_n}$ was the correct state. 
The same result holds for Bob, since at the crossing point Bob and Eve 
posses the same mixed state.

The mixture of the intermediate states may be interpreted as if Eve with 
probability $1/N$ has the state $\ket{{m}_{nj}}$ (there are N possible 
values of $j$). Eve, naturally, waits and performs her measurement after 
Alice has revealed in which basis the quNit was originally prepared. 
Then she measures her quNit in the same basis, which means that she uses 
either the basis $A$ or $A'$. 

Which means that the situation is the following: For the optimal 
eavesdropping 
strategy, Eve posses one of the intermediate 
states and she measures in one of the corresponding basis $A$ or $A'$. 
Whereas in the 
intercept/resent eavesdropping with the intermediate states, the situation 
is exactly the opposite namely, Eve has one 
of 
the basis states from $A$ or $A'$ and she measures the intermediate 
states. The two situations obviously lead to the same probabilities and 
hence the same amount of information.

\section{The Bell inequality in arbitrary dimension}
Recently there has been a big interest in generalizing various type of 
Bell inequalities \cite{bell}, \cite{bb}-\cite{bf} in higher dimension. 
The Bell inequality we present here \cite{BPG} makes use of the 
intermediate states, in a way similar to the CHSH inequality. 
This means that first we present the measurements and the quantum limit 
and only afterwards the local variable bound. So at first we just write 
down a particular sum of joint probabilities.

\subsection{The Bell inequality: The quantum mechanical limit}
Suppose that Alice and Bob share many maximally entangled state of two 
quNits. In 
the computational basis this state may be written as
\begin{eqnarray}
\ket{\psi}=\frac{1}{\sqrt{N}}\sum_{i=0}^{N-1}\ket{a_i,a_i}
\end{eqnarray}

For each of her quNits Alice has the choice of two measurements, namely 
to measure the basis $A$ or the basis $A'$. Whereas Bob for each of 
his quNits has the choice between $N^2$ binary measurements, corresponding 
to all the intermediate states of the two bases used by Alice.

In order to write down the Bell inequality, it is convenient to assign 
values to the various states. In the table below is shown the values:
\begin{center}
\begin{tabular}{|l||l|l||l|l|l|l|}\hline
value & $A$ & $A'$ & $M_0$ & $M_1$ & $\cdots$  & $M_{N-1}$\\
\hline \hline
0 & $\ket{a_0}$ &$\ket{{a_0}'}$ &$\ket{m_{00}}$& $\ket{m_{01}}$&$\cdots$
&
$\ket{m_{0,N-1}}$ \\\hline
1 & $\ket{a_1}$ &$\ket{{a_1}'}$ &$\ket{m_{11}}$& $\ket{m_{12}}$
&$\cdots$&
$\ket{m_{10}}$ \\\hline
$\cdots$&$\cdots$&$\cdots$&$\cdots$&$\cdots$&$\cdots$&$\cdots$\\
\hline
$N-1$ & $\ket{a_{N-1}}$ &$\ket{{a_{N-1}}'}$ &$\ket{m_{N-1,N-1}}$&
$\ket{m_{N-1,0}}$&$\cdots$&
$\ket{m_{N-1,N-2}}$ \\\hline
\end{tabular}
\end{center}
Notice that intermediate states have been organized into $N$ sets so that 
the value of the state is always given by the first index. Moreover this 
organization into the sets $M_0, \ldots, M_{N-1}$, simplifies the notation 
in what follows. However, it is important to remember that the states in 
each of the sets are {\bf not} orthogonal, in other words they do not form 
$N$ orthogonal bases.

The inequality is a sum of joint probabilities. And it is obtained by 
summing all the probabilities for when the results of the measurements 
are correlated and from this sum subtract all the probabilities when the 
results are not correlated, i.e.
\begin{eqnarray}
B_N&=&\sum p(\rm{results~correlated})\nonumber \\
&-&\sum
p(\rm{results~not~correlated})\nonumber
\end{eqnarray}

Suppose that Alice measures in the $A$ basis and Bob measures the 
projectors in the set $M_0$. For this combination of measurements, there 
are the following contributions to the sum $B_N$:
\begin{eqnarray}
P(M_0=A)&=&\sum_{i=0}^{N-1}p(m_{ii}\bigcap 
a_i)=\frac{1}{2}+\frac{1}{2\sqrt{N}}\\
P(M_0\neq A)&=&\sum_{i,j=0,j\neq i}^{N-1}p(m_{ii}\bigcap 
a_j)=\frac{1}{2}-\frac{1}{2\sqrt{N}}
\end{eqnarray}
Where $P(M_0=A)$ should be read as follows: Bob measures one of  the 
projectors in $M_0$ and Alice measures $A$, and Bob obtains the value 
which is correlated with Alice's result. On the other hand $P(M_0\neq A)$ 
means that Bob's result is not correlated with the result obtained by 
Alice. The probability $p(m_{kl}\bigcap a_n)=p(m_{kl}|a_n)p(a_n)$ is the 
joint probability for obtaining both $\ket{a_n}$ and $\ket{m_{kl}}$.

The same is the case if Bob measures the projectors in any of the other 
sets $M_1,\ldots, M_{N-1}$ and Alice always measures in $A$. And again if 
Bob uses $M_0$ and Alice $A'$. Which means we have 
$P(M_i=A)=P(M_0=A')=\frac{1}{2}+\frac{1}{2\sqrt{N}}$ and 
$P(M_i\neq A)=P(M_0\neq A')=\frac{1}{2}-\frac{1}{2\sqrt{N}}$. 

Now consider the case where Bob uses $M_1$ and Alice $A'$, in this case 
Bob consistently finds a value which is $N-1$ higher than the value which 
correlates him with Alice. To see this, assume for example that Bob has 
the state $\ket{a_{0}'}$ which is assigned the value $0$. But the state in 
$M_1$ which gives the correct identification of this state is 
$\ket{m_{N-1,0}}$, but this state has been assigned the value $N-1$. 
Similar for all the other states, which leads to 
$P(M_1=A'+(N-1))=\frac{1}{2}+\frac{1}{2\sqrt{N}}$ and 
$P(M_1\neq A'+(N-1))=\frac{1}{2}-\frac{1}{2\sqrt{N}}$.
Actually, when ever Alice measures $A'$ and Bob uses any of the $M_i$, Bob 
consistently finds a value which is $N-i$ higher than the one which 
correlates him with Alice. This means that 
$P(M_i=A'+(N-i))=\frac{1}{2}+\frac{1}{2\sqrt{N}}$ and   
$P(M_i\neq A'+(N-i))=\frac{1}{2}-\frac{1}{2\sqrt{N}}$. 

It is now possible to write and evaluate the sum $B_N$
\begin{eqnarray}
B_N&= &\sum_{i=0}^{N-1}P(M_i=A) -\sum_{i=0}^{N-1}P(M_i\neq A)\nonumber \\
&&+\sum_{i=0}^{N-1}P(M_i=A'+(N-i)) -\sum_{i=0}^{N-1}P(M_i\neq
A'+(N-i))\nonumber \\
&=&2N\left( \left(\frac{1}{2}+\frac{1}{2\sqrt{N}}\right)-
             \left(\frac{1}{2}-\frac{1}{2\sqrt{N}}\right) 
\right)\nonumber\\
&=& 2\sqrt{N}
\end{eqnarray}
The quantity $B_N$ is a sum of $2N\times N^2$ terms if written out 
explicitly. In the next section we show that a local variable model which 
tries to attribute definite values to the observables will reach a 
maximum value of 2. This shows that we have obtained a Bell inequality 
where the quantum violation grows with the squareroot of $N$.
\subsection{The Bell inequality: the local variable limit}
On Alice's side $a_0, \ldots, a_{N-1}$ are measured simultaneously in a 
single measurement as the basis $A$, which means that only one of them can 
come out true in a local variable model. The same is the case for 
${a}_{0}', \ldots, {a}_{N-1}'$ which is measured as the basis $A'$. This 
means that, for example, if $a_i$ is true, meaning that the measurement of 
$A$ will result in the outcome $a_i$, then all probabilities involving 
$a_j$ with $j\neq i$ must be zero. 
It is different on Bob's side where each $m_{kl}$ is measured 
independently 
and hence they may all be true at the same time in a local variable model.

Assume now that according to some local variable model $a_i$ and 
${a}_{j}'$ are true. At the same time, in principle, all the $m_{kl}$ 
could be true too. But notice now that the only $m$-state which will give 
a positive contribution to the quantity $B_N$ is the one which 
identifies 
both $a_i$ and ${a}_{j}'$ correctly, i.e. $m_{ij}$. This will give rise to 
a contribution of $+2$. Whereas  $m_{il}$ and $m_{kj}$ where only one
index is correct, will only identify of the states correctly and the other 
one wrong. This means that these states, since this gives rise to one 
correct and one wrong identification, will result in a zero contribution.
And finally the states $m_{kl}$ where both indices are wrong will only 
give rise to errors and will hence give a negative contribution of $-2$ 
to the sum $B_N$. Which means that
\begin{equation}
B_N\leq 2
\label{eq:bell}
\end{equation}
However, we have already shown that quantum mechanically it is possible to 
violate this limit. Quantum mechanically the limit is $2\sqrt{N}$. This 
means that we have obtained a Bell inequality where the violation 
increases with the squareroot of the dimension.

For N=3, the inequality has been checked in various ways numerically. 
First of all it has been checked that $2\sqrt{3}$ is indeed the quantum 
mechanical limit to this sum of probabilities and that this maximum 
is reached for the maximally entangled state. Moreover, it has been 
checked using "polytope software" \cite{bb,poly} that the inequality 
eq.(\ref{eq:bell}) is optimal for the measurement settings which we have 
presented here.

\section{Bell parameter as a function of $\rho_B$}
In this section we investigate how the Bell violation decreases as a 
function of the disturbance introduced by the eavesdropper. It is not 
necessary to think of it in terms of quantum cryptography and 
eavesdropping, but simply that the quantum channel from Alice to Bob is 
noisy and that the noise which is introduced is identical to the noise an 
eavesdropper using the optimal cloning machine would introduce. 

Assume, without loss of generality, that without disturbance,  Bob would have 
received the state $\ket{a_0}$, then we know that the mixed state that he 
obtains as a function of the disturbance can be written 
eq.(\ref{eq:rhobob})
\begin{eqnarray}
\rho_B=F_B\ket{{a}_{0}}\bra{{a}_{0}}+\frac{D_B}{N-1}\sum_{i=1}^{N-1}
\ket{{a}_{i}}\bra{{a}_{i}} \nonumber
\end{eqnarray}
In order to compute $S(\rho_B)$ it is enough to consider the case where 
Bob for example use the states in $M_0$ for his measurements, the rest of 
the terms in the inequality follows by symmetry.

All the states in the $M_0$ set are of the form $\ket{m_{jj}}$. First 
computing the various probabilities $\bra{m_{jj}}\rho_B\ket{m_{jj}}$, we 
find
\begin{eqnarray}
\bra{m_{jj}}\rho_B\ket{m_{jj}}&=&F_B 
\overbrace{\braket{m_{jj}}{a_0}\braket{a_0}{m_{jj}}}^{p(m_{jj}|a_0)} 
\nonumber \\
&+& \frac{D_B}{N-1}\sum_{i=1}^{N-1} 
\underbrace{\braket{m_{jj}}{a_i}\braket{a_i}{m_{jj}}}_{p(m_{jj}|a_i)}
\end{eqnarray}
There are two different cases which have to be checked independently, 
namely $j=0$ and $j\neq0$: For $j=0$ we have
\begin{eqnarray}
\bra{m_{00}}\rho_B\ket{m_{00}}=F_BF
+ (N-1)\frac{D_B}{N-1} \frac{D}{N-1}
\end{eqnarray}  
and for $j\neq 0$ we have
\begin{eqnarray}
{\bra{m_{jj}}\rho_B\ket{m_{jj}}}_{j\neq 0}=F_B\frac{D}{N-1}+
F\frac{D_B}{N-1} + (N-2)\frac{D_B}{N-1} \frac{D}{N-1}
\end{eqnarray}
Where we have used that $p(m_{00}|a_0)=F$ and $p{(m_{jj}|a_0)}_{j\neq 
0}=\frac{D}{N-1}$, see eq.(\ref{eq:psuc}) and eq.(\ref{eq:perror}).

In the inequality $\bra{m_{00}}\rho_B\ket{m_{00}}$ appears with a plus 
sign, since this is the probability of correctly identifying the state. 
At the same time ${\bra{m_{jj}}\rho_B\ket{m_{jj}}}_{j\neq 0}$ appears 
$N-1$ times with a minus sign in the inequality, since these correspond to 
all 
the possible errors. This means that we can define 
\begin{eqnarray}
s(\rho)&=&\bra{m_{00}}\rho_B\ket{m_{00}}-(N-1)
{\bra{m_{jj}}\rho_B\ket{m_{jj}}}_{j\neq 0} \nonumber \\
&=& F_B(F-D)-FD_B-\frac{N-3}{N-1}DD_B
\end{eqnarray}
There are $2N^2$ terms equal to $s(\rho_B)$ in the Bell 
inequality, since there are $N^2$ intermediate states and each of them 
appear twice (once for each basis $A$ and $A'$). On the 
other hand in each basis $A$ and $A'$ each state 
appears with a probability $1/N$, this mean that the total Bell parameter 
is equal to
\begin{equation}
S(\rho_B)=2Ns(\rho_B)=2N\left(  F_B(F-D)-FD_B-\frac{N-3}{N-1}DD_B \right)
\end{equation}

It is now possible to answer a very interesting question, namely for which 
disturbance is $S(\rho_B)=2$. Using the values of $F$ eq.(\ref{eq:psuc}) 
and 
$D$ eq.(\ref{eq:perror}) and 
expressing $F_B=1-D_B$, one find that $S(\rho_B)=2$ for
\begin{eqnarray}
D_{B}^{S=2}=\frac{N^{\frac{3}{2}}-\sqrt{N}-N+1}{ 
N^{\frac{3}{2}}+N^2 -2N}
\label{eq:DB}
\end{eqnarray}
This can be compared to the disturbance $D$ at the crossing point between 
the information lines, this is shown in figure 1. We find that it is 
only for $N=2$ that $D_{B}^{S=2}=D$, and hence only in two dimension that 
the inequality we have presented here can be used as a security measure in 
quantum cryptography. However, it should be stressed that the violation of 
the inequality stops before the crossing point is reached. So a violation 
of the inequality, in any dimension, still means that Alice and Bob are
within the secure zone. 

\samepage{
\begin{figure}[t]
\begin{center}   
\leavevmode
\hbox{%
\epsfxsize=5in
\epsffile{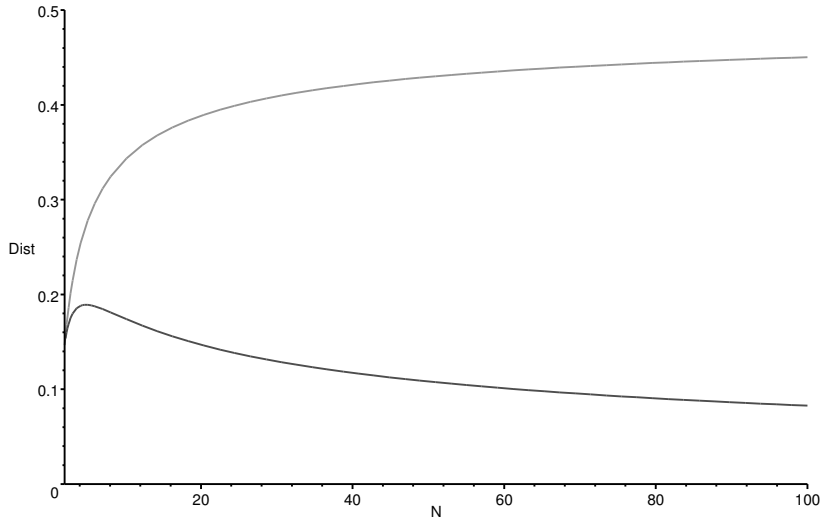}}
\caption{This figure shows as a function of the dimension, the disturbance 
at the crossing point $D$ (upper curve), and the disturbance for which the 
Bell violation stops $D^{S=2}$ (lower curve)} 
\end{center}  
\label{fig:fig1}
\end{figure}
}

\section{Interesting features of the $B_3$ inequality}
In this section we restrict ourself to three dimension in order to show in 
a simple way some interesting properties of the inequality. 
\subsection{Complex versus real numbers}
The first is 
related to the use of complex numbers. In the CHSH inequality for qubits, 
the maximal violation may be obtained by using real numbers only. Also the 
CGLMP inequality \cite{bf} show no difference between real and complex 
numbers. Here 
we show that if restricted to real numbers it is not possible to obtain 
maximal violation for the inequality we have presented. 

Numerically we have found the settings which leads to the largest 
violation when 
restricted to real numbers. On Alice's side the first basis is again the 
computational basis $A$, whereas the second basis $A^r$ (r stands 
for real) is found by making a $\pi/3$ 
rotation around the $(1,1,1)$ axis in $\Re^3$  and is explicitly given by 
\begin{eqnarray}
&&\ket{{a}_{0}^{r}}=
   \frac{1}{3}(2\ket{{a}_{0}}+2\ket{{a}_{1}}-1\ket{{a}_{0}})\nonumber\\
&&\ket{{a}_{1}^{r}}=
   \frac{1}{3}(-1\ket{{a}_{0}}+2\ket{{a}_{1}}+2\ket{{a}_{0}})\\
&&\ket{{a}_{2}^{r}}=
   \frac{1}{3}(2\ket{{a}_{0}}-1\ket{{a}_{1}}+2\ket{{a}_{0}})\nonumber
\end{eqnarray}
The intermediate states are defined in the same way, and the three sets 
$M_0$, $M_1$ and $M_2$ again consists of nonorthogonal states. We find the 
following probabilities:
\begin{eqnarray}
P(M_0=A)=5/6~~~~&,&~~~~P(M_0\neq A)=1/6 \nonumber \\
P(M_1=A)=4/6~~~~&,&~~~~P(M_1\neq A)=2/6 \nonumber \\
P(M_2=A)=5/6~~~~&,&~~~~P(M_2\neq A)=1/6 \nonumber \\
P(M_0=A^r)=5/6~~~~&,&~~~~P(M_0\neq A^r)=1/6 \nonumber \\
P(M_1=A^r +2)=4/6~~~~&,&~~~~P(M_1\neq A^r +2)=2/6 \nonumber \\
P(M_2=A^r +1)=5/6~~~~&,&~~~~P(M_2\neq A^r +1)=1/6 \nonumber 
\end{eqnarray}
Alice and Bob are still assumed to share the maximally entangled state 
$\ket{\psi}$. Inserting these probabilities in the $B_3$ inequality leads 
to $B_3=10/3\approx 3.333$, which is smaller than the maximal violation 
which is $2\sqrt{3}\approx 3.464$.

The explanation for this difference can be found in the fact that the 
inequality $B_N$ has been optimized for mutually unbiased bases. In two 
dimension it is possible to have two such bases, for example the $z$ and 
the $x-$basis are mutually unbiased and both real. But when going to 
higher dimension it is not the case, for example in three dimension it is 
not possible to have two mutually unbiased bases and have them both real. 
Which means that in order to reach the maximum value, for the inequality 
we have presented here, it is necessary to 
introduce complex numbers. Whereas the CGLMP inequality for qutrits has 
not 
been optimized for mutually unbiased bases, this explains why it does not 
require complex numbers. 

\subsection{Binary measurements versus basis measurements}
The $B_N$ inequality is on Bob's side optimized for the $N^2$ binary 
measurements corresponding to the intermediate states of the two bases 
chosen by Alice. However it is possible to impose the additional 
requirement that not only must the measurements chosen by Bob maximize 
the probabilities, but they must also form basis. In other words it is 
possible to require that the $M$-sets correspond to orthogonal bases 
$M^b$ (b refers to basis). 
We have considered this question in three dimensions. 

The $M^b$-basis which 
provide the optimal solution are defined in the following way: For the two 
mutually unbiased bases chosen by Alice there exist unitary operators 
$U_i$
such that
\begin{eqnarray}
U\ket{{a}_{i}}=\ket{{a}_{i}'}~~~,~~~U=A'\cdot A^{-1}
\end{eqnarray}
In this way the intermediate {\bf basis} is defined as 
\begin{eqnarray}
\ket{{m}_{ii}^{b}}=\sqrt{U}\ket{{a}_{i}}~~~,~~~M_i=\sqrt{U}A
\end{eqnarray}
Since $U$ is unitary $\sqrt{U}$ is well defined. It is possible to 
construct all three basis $M_{0}^{b}$, $M_{1}^{b}$ and $M_{2}^{b}$ in this 
way, choosing the unitary operator such that it transform the states in 
$A$ into any of the states in $A'$. This definition leads to the following 
probabilities
\begin{eqnarray}  
P({M}_{0}^{b}=A)=7/9~~~~&,&~~~~P({M}_{0}^{b}\neq A)=2/9 \nonumber \\
P({M}_{1}^{b}=A)=7/9~~~~&,&~~~~P({M}_{1}^{b}\neq A)=2/9 \nonumber \\
P({M}_{2}^{b}=A)=7/9~~~~&,&~~~~P({M}_{2}^{b}\neq A)=2/9 \nonumber \\
P({M}_{0}^{b}=A')=7/9~~~~&,&~~~~P({M}_{0}^{b}\neq A')=2/9 \nonumber \\
P({M}_{1}^{b}=A'+2)=7/9~~~~&,&~~~~P({M}_{1}^{b}\neq A'+2)=2/9 \nonumber \\
P({M}_{2}^{b}=A'+1)=7/9~~~~&,&~~~~P({M}_{2}^{b}\neq A'+1)=2/9 \nonumber
\end{eqnarray}
These probabilities may again be used in the $B_3$ inequality. But it is 
important to realize that evenif the notation for the inequality is the 
same, the interpretation is different. Since the states in 
the ${M}_{i}^{b}$-sets are orthogonal and ${M}_{i}^{b}$ are bases, Bob no 
longer chooses between $9$ different binary measurements but between three 
basis measurements. However, it is possible to check that the local 
variable limit is not changed, i.e. it is still $2$. Inserting the above 
probabilities leads to ${B}_{3}^{b}= 6(7/9 - 2/9)=10/3\approx 3.333$.

However, using basis measurements on Bob's side leads to some other 
interesting results. It turns out that it is possible to reduce the number 
of terms in the inequality. The $B_3$ inequality is the sum of all 
correct guesses, 
subtracting all the errors. Using the intermediate bases ${M}_{i}^{b}$ it 
is possible to subtract only half of the errors and in this way obtain 
a different inequality with a different local variable limit, namely 
$S_{12}\leq 3$, 
\begin{eqnarray} 
S_{12}&=&P({M}_{0}^{b}=A)+P({M}_{1}^{b}=A)+P({M}_{2}^{b}=A)\nonumber \\
&+&P({M}_{0}^{b}=A')+P({M}_{1}^{b}=A'+2)+P({M}_{2}^{b}=A'+1)\nonumber \\
&-&P({M}_{0}^{b}=A+1)-P({M}_{1}^{b}=A+1)-P({M}_{2}^{b}=A+1)\nonumber \\
&-&P({M}_{0}^{b}=A'+2)-P({M}_{1}^{b}=A'+1)-P({M}_{2}^{b}=A')\nonumber \\
&\leq&3
\end{eqnarray}
Inserting the above probabilities leads to the quantum 
mechanical maximum for this inequality namely, $S_{12}=6(7/9 -1/9)=4$.

\section{Resistance to noise}
In the recent papers on Bell inequalities, the strength of the inequality 
has been measured in terms of its resistance to noise \cite{zz,be,bf}. 
The question is how 
much noise can be added to the maximally entangled state, $\ket{\psi}$, 
and still obtain the Bell violation. The more noise which can be added to 
the system the better, since this means that the inequality is robust 
against noise. 

What is meant by noise naturally has to be specified. In the previous 
section we, for example, considered the noise which is introduced by an 
eavesdropper when she uses the optimally eavesdropping strategy. However, 
the noise which was until recently used in the measure of the strength of 
an inequality was uncolored noise. This means that the maximally entangled 
state is mixed with the maximally mixed state, so that the quantum state 
becomes
\begin{eqnarray}
\rho_{mix}=\lambda_{mix}\ket{\psi}\bra{\psi}+(1-\lambda_{mix})
\frac{\openone}{N^2}
\end{eqnarray}
This can be interpreted as if Bob with probability $\lambda_{mix}$ 
receives the maximally entangled state and with probability 
$1-\lambda_{mix}$ he receives the maximally mixed state. For the maximally 
entangled state the Bell inequality, $B_N$ (\ref{eq:bell}) has maximal 
violation, 
i.e. $S=2\sqrt{N}$. Whereas for the maximally mixed state each of the 
 probabilities in the inequality is equal to $1/N$, hence
$S(\openone/N^2)=2(2-N)$. This leads to
\begin{eqnarray}
S(\rho_{mix})=2 ~~~\Longleftrightarrow~~~ 
\lambda_{mix}^{B_N}=\frac{N-1}{N+\sqrt{N}-2}
\end{eqnarray}
For $N=3$, this is $\lambda_{mix}^{B_3}=\frac{2}{1+\sqrt{3}}\simeq 0.73$. 
In comparison, the 
CGLMP inequality  is more robust to this kind of noise, 
since they find a violation until $\lambda_{mix}^{CGLMP}\simeq 0.69$. 

Recently it has been argued that the use of uncolored noise in this 
measure lead to problems \cite{cp,bm}. At the same time a different kind 
of 
noise was 
introduced, namely to mix the maximally entangled state with the closest 
separable state, i.e.
\begin{eqnarray}
\rho_{cs}=\lambda_{sep}\ket{\psi}\bra{\psi}+(1-\lambda_{sep})\rho_{sep}
\end{eqnarray}
where $\rho_{sep}=\frac{1}{N}\sum_{i=0}^{N-1}\ket{a_i , a_i}\bra{a_i, 
a_i}$ \cite{PV}. Examining what happens to the Bell violation when 
introducing the 
state $\rho_{sep}$ in $B_N$ eq.(\ref{eq:bell}), shows that when Alice 
measures in the $A$ basis, Alice and Bob remain perfectly correlated - 
which means maximal violation of that part of the inequality concerning 
measurement combinations involving $A$.  
Whereas when Alice measures in the $A'$ basis Bob is left with the 
maximally mixed state, which means that all the joint probabilities 
involving using $A'$ on Alice's side are equal to $1/N$. In total this 
leads to
\begin{eqnarray}
S(\rho_{cs})=2 ~~~\Longleftrightarrow~~~
\lambda_{sep}^{B_N}=\frac{N-\sqrt{N}}{N+\sqrt{N}-2}
\end{eqnarray}
which for $N=3$ is 
$\lambda_{sep}^{B_3}=\frac{3-\sqrt{3}}{1+\sqrt{3}}\simeq 0.46$. Whereas 
the CGLMP inequality again has $\lambda_{sep}^{CGLMP}\simeq 0.69$. Which 
means that the inequality we have introduced here 
is much more robust to this kind of noise.

It should be stressed that the same measurement settings have been used in 
both 
evaluations of $\lambda$, and that the CGLMP inequality has been optimized 
to be resistant to the uncolored noise. It is nevertheless interesting 
to see how the robustness of the $B_N$ inequality change depending on the 
different noise added to the system.

\section{Minimum detection efficiency}

To conclude the study of the inequality (\ref{eq:bell}), let us consider
the minimum detection efficiency required to violate it. This question is
interesting both from a fundamental point of view (the so called
detector efficiency loophole \cite{detloophole}) and for the practical
question: How to test a quantum device, like a quantum cryptography
system? \cite{mayers98,qcrmp}. For simplicity we assume that all detectors
have the same efficiency $\eta$. The problem is what to do with the cases
that only one detector fires. A natural possibility
attributes the value zero to bob whenever his detector did not fire and a random
value to Alice whenever her detector did not fire. In this way, if only
Alice detects a quNit, the Bell function vanishes. Whereas, if only Bob detects, the
Bell function is the same as for the maximally mixed state, i.e. $2(2-N)$ as in
the previous section. Thus the inequality reads:
\beq
\frac{\eta^2\cdot 2\sqrt{N} + \eta(1-\eta)\cdot(0+2(2-N))}{\eta^2+2\eta(1-\eta)} \le 2
\eeq
From this inequality one finds the threshold efficiency:
\beq
\eta_{threshold}=\frac{N}{N+\sqrt{N}-1}
\eeq
For qubits, i.e. N=2, one recovers the welknown threshold, usually derived
from the Clauser-Horn inequality \cite{CH74}:
$\eta_{threshold}^{(N=2)}\approx 82.8\%$. This threshold is minimal and slightly better for N=4: 
$\eta_{threshold}^{(N=4)}=80\%$. For higher dimensions the
threshold increases and tends to 1.

It would be interesting to investigate the behavior of non-maximally
entangled states, since Eberhard found that for qubits the threshold then
decreases \cite{Eberhard93}. Let us mention that recentyly S. Massar proved that there are inequalities for which the threshold
tends to zero exponentially, at least for very large dimensions
\cite{massar} and, with colleagues he investigated situation similar to the one
studied in this section \cite{Zoology}.

\section{Conclusion}
For qubits the intermediate states play fundamental roles in at least 
three different place: intercept/resend eavesdropping in the BB84 
protocol for quantum cryptography, optimal eavesdropping also in the BB84 
protocol and 
in the CHSH-inequality for two entangled qubits. The work we have 
presented here, 
is the result of a study, of the use of these intermediate state in the 
same situations but in arbitrary dimension.

In this paper we have first discussed the generalization of the 
intermediates states of two mutually unbiased bases, showing that these 
states are in general not orthogonal and hence do not form basis as in the 
case for qubits. We have also discussed how they, nevertheless, can be 
use 
as binary measurements. With these measurements we have considered the 
same 
situations as known from the qubit case. 

We have considered eavesdropping in the generalized BB84 protocol (always 
considering only two bases). When the eavesdropper use the optimal 
eavesdropping strategy, her information increase as a function of the 
disturbance that she introduce and at the same time Bob's information is a 
decreasing function of the disturbance. For a given disturbance their 
information lines cross. We have shown that the amount of information that 
the eavesdropper obtain at this crossing point, is exactly the same amount 
of information she would have obtained using the simple intercept/resend 
strategy using the intermediate states. However leading to a much 
higher disturbance. This is explained by the fact that 
in any dimension Eve's mixed state can at the crossing point be decomposed 
into a sum of some of the intermediate states. Hence, in the optimal 
eavesdropping strategy, at the crossing point, Eve has one of the 
intermediate state but perform her measurement in the same basis as the 
state was originally prepared. Whereas in the intercept/resend strategy 
using the intermediate states as binary measurements, Eve has the state 
which was originally prepared by Alice, but measure one of the intermediate 
states. This means that the two situations are exactly opposite, and 
therefore lead to the same probabilities and hence the same information.

The maximal settings for the CHSH-inequality for qubits are two mutually 
unbiased bases on Alice's side and using the intermediate states on Bob's 
side. In the case of qubits the four intermediate states form two bases. 
This means that in this case both Alice and Bob have the choice of 
measuring one of two mutually unbiased bases. 

In higher dimension where the intermediate states do not form bases, Bob 
instead use the corresponding projectors as binary measurements. Which 
means that he choose between $N^2$ mutually incompatible measurements, 
whereas Alice still choose between two basis measurements. The 
generalized inequality we present has the local variable limit equal to 
$2$ in any dimension whereas the maximal quantum mechanical value is 
$2\sqrt{N}$. In other words we find a violation which increase with the 
squareroot of the dimension. Due to the construction we also obtain the 
familiar CHSH-inequality for $N=2$. 

It is known that the CHSH inequality may be used as a security measure in 
quantum cryptography for qubits. Since in this case a violation of the 
inequality is obtained until the disturbance introduced by the 
eavesdropper, reaches the disturbance at the crossing point of the 
information lines between Eve and Bob. Until this point Alice and Bob can 
use the fact that they share more mutual information than with Eve to 
obtain a secret key by means of one way privacy amplification. We have 
investigated the violation of the inequality we present here as a function 
for the disturbance introduced by the eavesdropper. We found that it is 
only for $N=2$ that the inequality can be used as a security measure, in 
the sense that in higher dimension the violation stops for a lower 
disturbance than the disturbance at the crossing point. This, however, 
does not mean that such an inequality does  not exist, it only shows that 
the inequality which mimics the situation from two dimension is not the 
one which has this property in higher dimension.

On the other hand the inequality we have presented here may stand as a 
result by itself and as a Bell inequality in arbitrary dimension is has 
many nice properties. First of all, compared to the other inequalities 
which have been presented recently, this inequality gives maximal 
violation for maximally entangled states. This means this inequality may 
be used as a measure of entanglements. Moreover we have shown in examples 
in three dimensions that this inequality require complex numbers in order 
to have maximal violation. Restriction to the use of real numbers lead to 
a smaller violation. In comparison, the CHSH inequality for qubits and the 
CGLMP inequality for qutrits show no difference between using real or 
complex numbers. The explanation is due to the fact that the inequality we 
present here is optimized for mutually unbiased bases and in three 
dimension it is not possible to have two such bases without the use of 
complex numbers. Whereas in two dimension the $x$ and $z-$ bases are 
mutually unbiased and both real, and for the CGLMP inequality the 
explanation is that it is not optimized of mutually unbiased bases.

We have also shown that imposing the additional constrain that the
$M$-sets actually form bases, leads to new inequalities. We have
explicitly given an example in three dimensions, showing the optimal
solution, for two basis measurements on Alice's side and three basis
measurements on Bob's side.

The strength of a Bell inequality has been measured in terms of 
its resistance to noise. Until recently the noise was taken to be 
uncolored noise, which means that the maximally entangled state is mixed 
with the maximally mixed state. The inequality we present here is less 
resistant to this kind of noise that other inequalities which have been 
presented recently. It should however be mentioned that these inequalities 
have been optimized for this kind of noise. However, recently it was 
argued that using the uncolored noise leads to problems. At the same time 
a different kind of noise was introduced, namely mixing the maximally 
entangled 
state with the closest separable state. When using this measure we find 
that the inequality we present here, is much more robust than, for 
example, 
the CGLMP-inequality.

\section*{Acknowledgments}
This work was done while H.B.-P. was at the Group of Applied Physics, 
University of Geneva, CH,  supported by the Danish 
National Science
Research Council (grant no. 9601645). This work was also supported by the 
Swiss NCCR "Quantum
Photonics" and by the European IST project EQUIP, sponsored by the Swiss 
OFES.

\end{document}